\documentclass[a4paper,12pt]{article}
\usepackage{amssymb}
\usepackage{amsmath}
\usepackage{mathtools}
\usepackage{slashed}
\usepackage{color}
\usepackage{indentfirst}
\usepackage{graphicx}
\usepackage{bbm}
\usepackage{csquotes} 
\usepackage{caption}
\usepackage{cite}
\newcommand{\drawsquare}[2]{\hbox{%
\rule{#2pt}{#1pt}\hskip-#2pt
\rule{#1pt}{#2pt}\hskip-#1pt
\rule[#1pt]{#1pt}{#2pt}}\rule[#1pt]{#2pt}{#2pt}\hskip-#2pt
\rule{#2pt}{#1pt}}

\newcommand{\fund}{\raisebox{-.5pt}{\drawsquare{6.5}{0.4}}}
\newcommand{\Ysymm}{\raisebox{-.5pt}{\drawsquare{6.5}{0.4}}\hskip-0.4pt%
        \raisebox{-.5pt}{\drawsquare{6.5}{0.4}}}
\newcommand{\Yasymm}{\raisebox{-3.5pt}{\drawsquare{6.5}{0.4}}\hskip-6.9pt%
        \raisebox{3pt}{\drawsquare{6.5}{0.4}}}
\newcommand{\antifund}{\overline{\fund}}

\renewcommand{\thefootnote}{\fnsymbol{footnote}}

\begin{document}

\title{
\begin{flushright}
\begin{minipage}{0.2\linewidth}
\normalsize
EPHOU-20-0001\\
KEK-TH-2187 \\*[50pt]
\end{minipage}
\end{flushright}
{\Large \bf 
Classification of discrete modular symmetries \\ in Type IIB flux vacua
\\*[20pt]}}

\author{Tatsuo Kobayashi$^{a}$\footnote{
E-mail address: kobayashi@particle.sci.hokudai.ac.jp
}
\ and\
Hajime~Otsuka$^{b}$\footnote{
E-mail address: hotsuka@post.kek.jp
}\\*[20pt]
$^a${\it \normalsize 
Department of Physics, Hokkaido University, Sapporo 060-0810, Japan} \\
$^b${\it \normalsize 
KEK Theory Center, Institute of Particle and Nuclear Studies, KEK,}\\
{\it \normalsize 1-1 Oho, Tsukuba, Ibaraki 305-0801, Japan}}
\maketitle

\date{
\centerline{\small \bf Abstract}
\begin{minipage}{0.9\linewidth}
\medskip 
\medskip 
\small
We classify  discrete modular symmetries 
in the effective action of Type IIB string on toroidal 
orientifolds with three-form fluxes, emphasizing on $T^6/\mathbb{Z}_2$ and 
$T^6/(\mathbb{Z}_2\times \mathbb{Z}_2^\prime)$ orientifold backgrounds.  
On the three-form flux background, the modular 
group is spontaneously broken down to its congruence subgroup whose pattern is 
severely constrained by a quantization of fluxes and tadpole cancellation conditions. 
We explicitly demonstrate that the congruence subgroups appearing 
in the effective action arise on magnetized D-branes wrapping certain cycles of tori. 
\end{minipage}
}


\renewcommand{\thefootnote}{\arabic{footnote}}
\setcounter{footnote}{0}
\thispagestyle{empty}
\clearpage
\addtocounter{page}{-1}

\tableofcontents

\section{Introduction}\label{sec:introduction}

In the low-energy effective action of higher-dimensional theory, moduli are ubiquitous fields and they have certain symmetries, originating from the higher-dimensional gauge and/or Lorentz symmetries. 
As an example, axion-like fields have the so-called shift symmetries associated with the gauge symmetries of higher-form fields, which play an important role in solving the strong CP \cite{Peccei:1977hh} and hierarchy problems~\cite{Graham:2015cka}, the candidates of dark matter~\cite{Preskill:1982cy,Abbott:1982af,Dine:1982ah} and  inflaton~\cite{Freese:1990rb,Silverstein:2008sg,McAllister:2008hb}. 
Such  shift symmetries are useful to control the action against the higher-order corrections.

Torus and orbifold compactifications have the modular symmetry.
For example, the two-dimensional tori and orbifolds have the modular symmetry,  $SL(2,\mathbb{Z})$.
Modular symmetries in effective supergravity theory have been studied, e.g. 
for moduli stabilization, supersymmetry breaking \cite{Ferrara:1990ei,Cvetic:1991qm} and inflation models \cite{Kobayashi:2016mzg}.
Moreover, Yukawa couplings and higher order couplings depend on moduli.
(See for moduli-dependent couplings in heterotic orbifold models 
\cite{Hamidi:1986vh}, intersecting D-brane models \cite{Cvetic:2003ch}, 
and magnetized D-brane models \cite{Cremades:2004wa}.)
They transform non-trivially under the modular symmetry.
Then, Yukawa matrices transform non-trivially, 
and the modular group transforms flavors non-trivially, that is, the flavor symmetry.
Indeed, matter fields on these compactifications transform each other under the modular group.
For example, a finite number of zero-modes appear on torus and orbifold compactifications with magnetic fluxes.
These zero-modes transform under the modular group.
That is, these zero-modes 
become finite representations of the modular group or its subgroups, which are finite discrete groups 
\cite{Kobayashi:2017dyu,Kobayashi:2018rad}.\footnote{See also for recent relevant works in magnetized D-brane models 
\cite{Kobayashi:2016ovu,Kobayashi:2018bff,Kariyazono:2019ehj}
and heterotic orbifold models \cite{Baur:2019kwi,Nilles:2020nnc}.}

Recently, discrete subgroups arising from the $SL(2,\mathbb{Z})$ modular group as well as its congruence subgroups have been utilized for the flavor symmetry of quarks/leptons including CP violation and applications for leptogenesis and dark matter candidates~\cite{Feruglio:2017spp,Kobayashi:2018vbk,Penedo:2018nmg,Criado:2018thu,Kobayashi:2018scp,Novichkov:2018ovf,
Novichkov:2018nkm,deAnda:2018ecu,Okada:2018yrn,Kobayashi:2018wkl,Novichkov:2018yse,Ding:2019xna,
Nomura:2019jxj,Novichkov:2019sqv,Okada:2019uoy,deMedeirosVarzielas:2019cyj,Nomura:2019yft,Kobayashi:2019rzp,
Liu:2019khw,Okada:2019xqk,Kobayashi:2019mna,Ding:2019zxk,Okada:2019mjf,King:2019vhv,Nomura:2019lnr,Okada:2019lzv,
Criado:2019tzk,Kobayashi:2019xvz,Asaka:2019vev,Gui-JunDing:2019wap,Zhang:2019ngf,Wang:2019ovr,Kobayashi:2019uyt,
Nomura:2019xsb,Kobayashi:2019gtp,Lu:2019vgm,Wang:2019xbo}.
The reason to attract attention of researchers is that the quotients $\Gamma_N=SL(2,\mathbb{Z})/\Gamma(N)$ 
by  congruence subgroups  $\Gamma(N)$ correspond to 
the well-known discrete finite groups such as 
$\Gamma_2\simeq S_3$, $\Gamma_3\simeq A_4$ $\Gamma_4\simeq S_4$, $\Gamma_5\simeq A_5$, which provide interesting flavor structures of quarks and/or leptons~\cite{deAdelhartToorop:2011re}.

The three-form fluxes can stabilize some or all of the complex structure moduli \cite{Giddings:2001yu}.
Then, the geometrical symmetries of $T^6=T^2_1\times T^2_2 \times T^2_3$ with 
non-vanishing three-form fluxes can be different from  $\bigotimes_{i=1}^3 SL(2,\mathbb{Z})_i$.
They would provide us with new possibilities for starting points toward the above studies of 
particle physics and cosmologies.

In this paper, we study a simple Type IIB flux compactification on toroidal orientifolds with and without D-branes, 
where the modular symmetries associated with tori are partially broken 
into subgroups by the three-form fluxes. 
Subgroups of the modular group  $\bigotimes_{i=1}^3 SL(2,\mathbb{Z})_i$ 
emerge in the flat directions of moduli fields such that the modular transformation is viable in the low-energy effective action. 
Our aim is to classify the modular groups and its subgroups in the effective action if it exists, meaning that we do not consider the stabilization of the remaining massless moduli in this work. Such an approach is similar to 
the recent swampland program~\cite{Vafa:2005ui,Ooguri:2006in,Palti:2019pca} where the allowable moduli space of massless modes is taken into account. 

This paper is organized as follows. 
In Section~\ref{sec:2}, we first show the realization of the modular symmetry and its breaking in the low-energy effective action of Type IIB string theory on a factorizable 6-torus subject to $\mathbb{Z}_2$ identification, i.e. $T^6/\mathbb{Z}_2$. 
The idea to obtain the congruence subgroups of the modular group is basically given in \cite{Hebecker:2017lxm}. They focus on the flat axionic direction which can be enlarged to the Planckian field region by an existence of fluxes. 
In this paper, we extend their analysis and classify the remaining modular symmetry in the low-energy effective action. 
Next, we move on to $T^6/(\mathbb{Z}_2\times \mathbb{Z}_2^\prime)$ orientifold background where several semi-realistic models are proposed in \cite{Blumenhagen:2003vr,Cascales:2003zp,Marchesano:2004xz,Cvetic:2005bn}. 
In Section~\ref{sec:3}, we classify the breaking pattern of the modular group 
on $T^6/(\mathbb{Z}_2\times \mathbb{Z}_2^\prime)$ orientifold with magnetized 
D-branes wrapping certain cycles of tori. 
Similar to the analysis in Section~\ref{sec:2}, 
we enumerate the possible congruence subgroups in a concrete three-generation 
model. 
It turns out that the remaining modular symmetry in the effective action is severely constrained by the quantization of fluxes and tadpole cancellation conditions. 
Section~\ref{sec:conclusion} is devoted to the conclusions and discussions. 
In the Appendix, we show our conventions of congruence subgroups.

\section{Modular symmetry on $T^6/\mathbb{Z}_2$ toroidal orientifold}
\label{sec:2}

In this section, we briefly review the modular symmetry in four-dimensional (4D) effective action of 
Type IIB string on $T^6/\mathbb{Z}_2$ toroidal orientifold with 
three-form fluxes. 
After demonstrating the breaking mechanism of the modular 
symmetry discussed in \cite{Hebecker:2017lxm}, we extend their analysis and 
classify patterns of 
congruence subgroups in the low-energy effective action.

\subsection{Effective action}
The setup is the 4D effective action of Type IIB string on 
toroidal orientifold, in particular the factorizable 6-torus 
$T^6/\mathbb{Z}_2 = (T_1^2 \times T_2^2\times T_3^2)/\mathbb{Z}_2$. 
The effective action of the closed string moduli on this background is 
described in the 4D ${\cal N}=1$ language, namely the K\"ahler potential $K$ and 
the superpotential $W$\footnote{We follow the notation of \cite{Kachru:2002he} and adopt the reduced 
Planck mass unit $M_{\rm Pl}=1$.},
\begin{align}
    K = -\ln (-i(\tau -\Bar{\tau})) -2\ln {\cal V} 
    -\ln(i(\tau_1-\Bar{\tau}_1)(\tau_2-\Bar{\tau}_2)(\tau_3-\Bar{\tau}_3)), 
    \label{eq:Keff}
\end{align}
where $\tau=C_0 +i e^{-\phi}$ is the axio-dilaton, $\tau_i$ with $i=1,2,3$ 
are three complex structure moduli on $T_i^2$ and ${\cal V}$ denotes the volume of torus. 
The superpotential is generated by background three-form fluxes $G_3 =F_3 -\tau H_3$~\cite{Gukov:1999ya}
\begin{align}
    W = \frac{1}{l_s^2}\int \Omega \wedge G_3,
    \label{eq:Weff}
\end{align}
where $l_s = 2\pi \sqrt{\alpha^\prime}$ and a holomorphic three-form 
$\Omega$ is specified by the six real coordinates on $T^6$ ($x^i,y^i)$ with $i=1,2,3$ 
and the complex structure moduli $\tau^i$,
\begin{align}
    \Omega &= dz_1 \wedge dz_2 \wedge dz_3,    
    \label{eq:Omega}
\end{align}
with $dz^i =dx^i +\tau^i dy^i$. 
In a similar way, $G_3$ can be expanded on the basis of $H^3(T^6,2\mathbb{Z})$, 
($\alpha_I$, $\beta^J$), satisfying $\int_{T^6}\alpha_I \wedge \beta^J =\delta^{J}_I$,
\begin{align}
\frac{1}{l_s^2}F_3 &= a^0 \alpha_0 +a^{i}\alpha_{i} +b_{i}\beta^{i} +b_0 \beta^0,
\nonumber\\
\frac{1}{l_s^2}H_3 &= c^0 \alpha_0 +c^{j}\alpha_{i} +d_{i}\beta^{i} +d_0 \beta^0,
\label{eq:F3H3_T6Z2}
\end{align}
where $a^{0,1,2,3},b_{0,1,2,3}$ and $c^{0,1,2,3},d_{0,1,2,3}$ are 
quantized to be even integers according to 
\begin{align}
    \frac{1}{l_s^2}\int F_3\in 2 \mathbb{Z},\qquad
    \frac{1}{l_s^2} \int H_3 \in 2 \mathbb{Z}.
    \label{eq:T6Z2quant}
\end{align}
In this paper, we restrict ourselves to consider even integers; otherwise, 
exotic O3-plane contributions are necessary in the system~\cite{Hanany:2000fq,Witten:1997bs}. 
The basis of three-form is explicitly given by
\begin{align}
    \alpha_0 &= dx^1 \wedge dx^2 \wedge dx^3, \qquad
    \alpha_{i} = \frac{1}{2}\epsilon_{ilm}dx^l \wedge dx^m \wedge dy^i\;\; (1 \leq i\leq 3),
\nonumber\\
    \beta^0 &= dy^1 \wedge dy^2 \wedge dy^3, \qquad
    \beta^{i} = -\frac{1}{2}\epsilon_{ilm}dy^l \wedge dy^m \wedge dx^i\;\; (1 \leq i\leq 3).    \label{eq:basis}
\end{align}

Then, we can explicitly write down the superpotential:
\begin{align}
    W &= (a^0 -\tau c^0)\tau_1\tau_2\tau_3 - (a^1 -\tau c^1)\tau_2\tau_3
    - (a^2 -\tau c^2)\tau_1\tau_3 - (a^3 -\tau c^3)\tau_1\tau_2
    \nonumber\\
    &- \sum_{i=1}^3 (b_i -\tau d_i)\tau_i - (b_0 -\tau d_0).
\label{eq:W}
\end{align}
Note that three-form fluxes induce the D3-brane charge $N_{\rm flux}=\int_{T^6} H_3 \wedge F_3$ which appears in the cancellation condition of 
D3-brane charge
\begin{align}
    N_{{\rm D}3} + \frac{1}{2}N_{\rm flux} =\frac{1}{4}N_{{\rm O}3}=16.
\end{align}
Here we employ the fact that there exist 64 O3-planes 
associated with  64 fixed points on $T^6/\mathbb{Z}_2$ orientifold background. 
Thus, the above tadpole cancellation condition is simplified as
\begin{align}
    c^0b_0 -d_0a^0 +\sum_i (c^ib_i-d_ia^i) =2(16- N_{{\rm D}3}) \leq 32,
    \label{eq:tadT6Z20}
\end{align}
where we do not consider the presence of anti D3-branes to 
preserve the supersymmetry.

\subsection{Modular symmetry of the effective action}
Before we explain the mechanism to derive the discrete modular groups, 
we briefly review the modular symmetry of the effective action, following~\cite{Hebecker:2017lxm}. 
The effective K\"ahler potential (\ref{eq:Keff}) and superpotential (\ref{eq:Weff}) are 
invariant under $SL(2,\mathbb{Z})_\tau$, where $SL(2,\mathbb{Z})_\tau$ denotes the modular 
group associated with the axio-dilaton. 
The explicit modular transformations for the axio-dilaton $\tau$ and 
the pair of three-form fluxes $(F_3, H_3)$ are given by
\begin{align}
\tau^\prime = R(\tau)=\frac{p\tau +q}{s\tau +t},\qquad
    \begin{pmatrix}
       F_3^\prime\\
       H_3^\prime
    \end{pmatrix}
    =     
    \begin{pmatrix}
       q & p\\
       t & s
    \end{pmatrix}    
     \begin{pmatrix}
       F_3\\
       H_3
    \end{pmatrix}   
    ,
\end{align}
where $p,q,s,t$ are integers constrained by $pt-qs=1$ and $R\in SL(2,\mathbb{Z})_\tau$.

Similarly,  each torus $T^2_i$ has the modular symmetry  $SL(2,\mathbb{Z})_i$ 
for vanishing three-form fluxes.
Each torus $T^2_i=R^2_i/\Lambda_i$ is defined 
by the two-dimensional Euclidean space $R^2_i$ modded out by the lattice $\Lambda_i$, 
and  the lattice $\Lambda_i$ is spanned by the basis vectors $(e_{x^i}, e_{y^i})$.
The same lattice is spanned by other basis vectors $(e_{x^i}^\prime, e_{y^i}^\prime)$ 
satisfying 
\begin{align}
    \begin{pmatrix}
       e_{y^i}^\prime\\
       e_{x^i}^\prime
    \end{pmatrix}
    =     R_i
     \begin{pmatrix}
       e_{y^i}\\
       e_{x^i}
    \end{pmatrix}   
    ,
\end{align}
with
\begin{align}
    R_i = 
    \begin{pmatrix}
     p_i & q_i \\
     s_i & t_i
    \end{pmatrix}
\in SL(2,\mathbb{Z})_i
    ,
\end{align} 
with $p_it_i -q_is_i =1$. 
This is the modular transformation, $SL(2,\mathbb{Z})_i$.
Then, the shape of each torus $T^2_i$, i.e., the modulus parameter $\tau_i$ transforms
\begin{align}
    \tau_i\equiv \frac{e_{y^i}}{e_{x^i}}\rightarrow 
    \tau_i^\prime\equiv \frac{e_{y^i}^\prime}{e_{x^i}^\prime} 
    = \frac{p_i \tau_i +q_i}{s_i \tau_i +t_i} =R_i(\tau_i),
\end{align}
under the modular transformation.
It is noted that the generators of the modular symmetry are given by
\begin{align}
    S\,:\,\tau_i \rightarrow -\frac{1}{\tau_i},\qquad
    T\,:\,\tau_i \rightarrow \tau_i+1,
\end{align}
which satisfy $S^2=(ST)^3=1$.

When we introduce the coordinates of each torus by choosing $e_{y^i}=\tau_i$ and $e_{x^i}=1$,
\begin{align}
    z_i = x_i + \tau_i y_i = 
    (y_i, x_i)\cdot 
    \begin{pmatrix}
     \tau_i \\
     1
    \end{pmatrix}
    ,
\end{align}
it is equivalent to
\begin{align}
    z_i^\prime = x_i^\prime + \tau_i^\prime y_i^\prime = 
    (y_i, x_i)R^{-1}_iR_i 
    \begin{pmatrix}
     \tau_i \\
     1
    \end{pmatrix}
    ,
\end{align}
under the modular transformation $R_i \in SL(2,\mathbb{Z})_i$. 
Hence, the coordinates of each torus $(y_i, x_i)$ are related 
to $(y_i^\prime, x_i^\prime)$,
\begin{align}
    \begin{pmatrix}
       y_i^\prime\\
       x_i^\prime
    \end{pmatrix}
    =     (R_i^{-1})^T
     \begin{pmatrix}
       y_i\\
       x_i
    \end{pmatrix}   
    .
\end{align}
It indicates that  under the modular transformations of $\bigotimes_{i=1}^3 SL(2,\mathbb{Z})_i$ associated with three 2-tori, 
not only the complex structure moduli $\tau_i$ are related to $\tau_i^\prime$, 
but also the three-form fluxes $(F_3, H_3)$ in Eq.~(\ref{eq:F3H3_T6Z2}) 
non-trivially transform to $(F_3^\prime, H_3^\prime)$ because of the coordinate transformations 
of $(y_i, x_i)$. 
Recalling that the K\"ahler potential of the moduli $\tau_i$ transforms
\begin{align}
    -\ln(i(\tau_1-\Bar{\tau}_1)(\tau_2-\Bar{\tau}_2)(\tau_3-\Bar{\tau}_3))
    \rightarrow
    -\ln(i(\tau_1-\Bar{\tau}_1)(\tau_2-\Bar{\tau}_2)(\tau_3-\Bar{\tau}_3)) +
    \ln \biggl[ \Pi_{i=1}^3 |s_i\tau_i +t_i|^2\biggl],
\end{align}
the effective action has a modular invariance only when the transformation of the superpotential~(\ref{eq:Weff}) is 
provided by 
\begin{align}
    W \rightarrow \frac{W}{\Pi_{i=1}^3(s_i\tau_i +t_i)}.
    \label{eq:Wtransf}
\end{align}
Note that it is enforced by the fact that 
the K\"ahler potential and the superpotential appear in the combination $K+\ln |W|^2$ 
in the 4D ${\cal N}=1$ supergravity action. 
Hence, taking into account the modular transformation of the holomorphic three-form
\begin{align}
    \Omega \rightarrow \frac{\Omega}{\Pi_{i=1}^3(s_i\tau_i +t_i)},
    \label{eq:Omegatrf}
\end{align}
$\bigotimes_{i=1}^3 SL(2,\mathbb{Z})_i$ modular symmetries remain in the effective action, 
only when $G_3$ itself is invariant under the modular transformations.

\subsection{Congruence subgroups of the modular group}

In this section, we discuss the congruence subgroups of the modular group, which 
can arise by the spontaneous breaking of the modular group on 
the three-form flux background. 

To demonstrate the realization of the congruence subgroups of the modular groups, 
we consider the following simplified superpotential
\begin{align}
    W &= \tau_3\biggl[ a^0 \tau_1\tau_2 - a^1\tau_2
    - a^2\tau_1 -b_3\biggl] 
    +\tau \biggl[ c^3\tau_1\tau_2 +d_1\tau_1 +d_2\tau_2 +d_0\biggl]
    \nonumber\\
     &= (\tau_3 -f\tau)\biggl[ a^0 \tau_1\tau_2 
    - a^1\tau_2 - a^2\tau_1 -b_3\biggl],
    \label{eq:Wgeneral}
\end{align}
where we chose the nonvanishing fluxes as
\begin{align}
    c^3= -fa^0,\qquad d_1= f a^2,\qquad d_2 =fa^1,\qquad d_0= fb_3,
\end{align}
and the other fluxes vanish. 
The supersymmetric minimum is given by
\begin{align}
     \tau_3 &=f\tau, \qquad 
     \tau_1 =\frac{a^1\tau_2 +b_3}{a^0\tau_2-a^2},
     \label{eq:vacuumGeneral}
\end{align}
on which there exist flat directions on the moduli spaces. 
In the following subsections, we explicitly show the 
possible congruence subgroups on such flat directions. Note that the above fluxes induce the D3-brane charge
\begin{align}
    N_{\rm flux} =-d_0a^0 +c^3b_3-d_1a^1-d_2a^2= -2d_0a^0-2d_1a^1,
\end{align}
which is quantized in multiples of 8 on $T^6/\mathbb{Z}_2$ background 
due to Eq.~(\ref{eq:T6Z2quant}).

\subsubsection{Model 1}
\label{sec:model1}

We first discuss the simplest case where the superpotential (\ref{eq:Wgeneral}) reduces to be 
\begin{align}
    W &= -(\tau_3 -f\tau)(a^2 \tau_1 +a^1\tau_2),
      \label{eq:Wsim}
\end{align}
by further setting $a^0=b_3=0$ in (\ref{eq:Wgeneral}). 
Note that the flux quanta $a^{1,2}$ are even integers; 
otherwise, we have to introduce exotic O3-plane contributions. 
The above form of the superpotential has already been discussed in \cite{Hebecker:2017lxm}. 
Such a superpotential is induced by the following set of three-form fluxes:
\begin{align}
    F_3 &= (a^2 dx_1 \wedge dy^2 - a^1 dy^1\wedge dx_2)\wedge dx_3 
    \equiv A_{ij} d\xi_1^i \wedge d\xi_2^j \wedge dx_3,
    \nonumber\\
    H_3 &= -f(a^2 dx_1 \wedge dy^2 + a^1 dy^1\wedge dx_2)\wedge dy_3,    
    \equiv -f A_{ij} d\xi_1^i \wedge d\xi_2^j \wedge dy_3,
\end{align}
with 
\begin{align}
    A_{ij} \equiv 
     \begin{pmatrix}
       0 & a^1\\
       a^2 & 0
    \end{pmatrix}       
,\qquad 
    \xi_i \equiv 
     \begin{pmatrix}
       y_i\\
       x_i
    \end{pmatrix}       
.
\end{align}
The supersymmetric minimum $\partial_\tau W = \partial_{\tau_i} W=W=0$ with $i=1,2,3$ has two flat directions,
\begin{align}
    \tau_3 = f\tau,\qquad
    a^2 \tau_1 =-a^1\tau_2,
\end{align}
from which we require ${\rm sign}(f)=-{\rm sign}(a^1a^2)=1$ to realize ${\rm Im}(\tau),{\rm Im}(\tau_{1,2,3})>0$ in our conventions. 
Then, the tadpole cancellation condition (\ref{eq:tadT6Z20}) reads
\begin{align}
    0\leq -fa^1a^2 \leq 16,
    \label{eq:tadT6Z2}
\end{align}
from which $f\neq 1$ gives the severe condition to the choice of $a^1$ and $a^2$. 
As discussed later, only the ratio $a^1/a^2$ is important to classify the modular subgroups in the 
flat direction $a^2\tau_1 =-a^1\tau_2$. 
Thus the tadpole cancellation condition (\ref{eq:tadT6Z2}) with 
$f\neq 1$ restricts the choice of $a^1$ and $a^2$ in comparison with the $f=1$ case, 
which has the least constraint on $a^1$ and $a^2$. 
For that reason, we restrict ourselves to the $f=1$ case in the following discussion.

We first focus on the $SL(2,\mathbb{Z})_3$ modular transformations of the $(F_3, H_3)$ pair provided by
\begin{align}
    \begin{pmatrix}
       F_3^\prime\\
       H_3^\prime
    \end{pmatrix}
    =   R_3 
     \begin{pmatrix}
       F_3\\
       H_3
    \end{pmatrix}   
    ,
\end{align}
from which the total action is not invariant under $SL(2,\mathbb{Z})_3$. 
This is because $G_3$ itself transforms under the modular transformation 
which does not lead to the transformation of the superpotential as in (\ref{eq:Wtransf}). 
However, if we identify $R_3=R$ and take $f=1$, a diagonal part of $SL(2,\mathbb{Z})_\tau \times SL(2,\mathbb{Z})_3$ 
remains in the effective action. 
Indeed, in such a case, we can verify that the vacuum condition $\tau_3 = \tau$ still holds after the 
modular transformation,
\begin{align}
    \tau_3^\prime =R_3(\tau_3) =R(\tau_3) = R(\tau)= \tau^\prime,
    \label{eq:tautau3}
\end{align}
where we employ $R_3=R$ and $\tau_3 = \tau$.

Next, let us discuss the modular transformations on $T_1^2\times T_2^2$, 
namely $\tau_1^\prime =R_1 (\tau_1)$ and $\tau_2^\prime =R_2 (\tau_2)$, 
under which three-form fluxes transform
\begin{align}
F_3 &\rightarrow (R_1^{-1} A (R_2^{-1})^T)_{ij} d\xi_1^i \wedge d\xi_2^j \wedge dx_3,
\nonumber\\
H_3 &\rightarrow -f(R_1^{-1} A (R_2^{-1})^T)_{ij} d\xi_1^i \wedge d\xi_2^j \wedge dy_3.
\label{eq:F3H3}
\end{align}
To be invariant under the modular transformations, we require that 
\begin{align}
    R_1^{-1} A (R_2^{-1})^T=A,
\end{align}
which is possible under
\begin{align}
    R_2 = A^T (R_1^{-1})^T (A^{-1})^T = 
    \begin{pmatrix}
       p_1 & -q_1a^2/a^1 \\
       -s_1 a^1/a^2 & t_1
    \end{pmatrix}
    .
    \label{eq:R2model1}
\end{align}    
It turns out that $q_1$ and $s_1$ are multiples of $a^1$ and $a^2$, respectively. 
Similar to $\tau_3 = \tau$, we can  check that 
$a^2\tau_1 = -a^1\tau_2$ also holds after the modular transformations:
\begin{align}
    a^2\tau_1^\prime =a^2R_1(\tau_1) =a^2\frac{p_1\tau_1+q_1}{s_1\tau_1 +t_1}
    =-a^1\frac{p_1\tau_2-q_1a^2/a^1}{-(s_1a^1/a^2)\tau_2 +t_1} =-a^1R_2(\tau_2) =-a^1\tau_2^\prime,
    \label{eq:tau1tau2}
\end{align}
where we employ $a^2\tau_1= -a^1\tau_2$.

Interestingly, the fundamental region of the moduli spaces reduces to the region characterized by the matrices $R_1$ and $R_2$, enumerated as follows:
\begin{itemize}
\item $|a^2/a^1| = n\,(\in \mathbb{Z}_{>0})$
\begin{align}
    R_1 =
    \begin{pmatrix}
       1 & 1 \\
       0\,({\rm mod}n) & 1
    \end{pmatrix}
    \in 
    \Gamma_0(n),
    \qquad
    R_2 =
    \begin{pmatrix}
       1 & 0\,({\rm mod}n) \\
       1 & 1
    \end{pmatrix}
    \in 
    \Gamma^0(n),
\label{eq:Congruence1}
\end{align}
\item $|a^1/a^2| = n\,(\in \mathbb{Z}_{>0})$
\begin{align}
    R_1 =
    \begin{pmatrix}
       1 & 0\,({\rm mod}n) \\
       1 & 1
    \end{pmatrix}
    \in 
    \Gamma^0(n),
    \qquad
    R_2 =
    \begin{pmatrix}
       1 & 1 \\
       0\,({\rm mod}n) & 1
    \end{pmatrix}
    \in 
    \Gamma_0(n),
\label{eq:Congruence2}
\end{align}
\item gcd$(a^1, a^2) = 1$
\begin{align}
    R_1 =
    \begin{pmatrix}
       1 & 0\,({\rm mod}a^1) \\
       0\,({\rm mod}a^2) & 1
    \end{pmatrix}
    \in 
    \Gamma(|a^1a^2|)
,
    R_2 =
    \begin{pmatrix}
       1 & 0\,({\rm mod}a^2) \\
       0\,({\rm mod}a^1) & 1
    \end{pmatrix}
    \in 
    \Gamma(|a^1a^2|).
\label{eq:Congruence3}
\end{align}
\end{itemize}
Several congruence subgroups are defined in Appendix~\ref{app}. 
Note that $\Gamma_0(1)=\Gamma^0(1)=\Gamma(1)=SL(2,\mathbb{Z})$.
It indicates that the $SL(2, \mathbb{Z})_{1,2}$ groups are 
spontaneously broken to their congruence subgroups $\Gamma^{(1)}$ and $\Gamma^{(2)}$. 
Given that $a^1$ and $a^2$ are quantized to be even on $T^6/\mathbb{Z}_2$ 
background, we obtain the subgroups of $SL(2, \mathbb{Z})_{1,2}$ 
as shown in Table~\ref{tab:T6Z2}. 

\begin{table}[htb]
    \centering
    \begin{tabular}{|c|c|} \hline
   $f$    & $\{\Gamma^{(1)}, \Gamma^{(2)}\}$\\ \hline \hline
   1    & $\{\Gamma_0(2), \Gamma^0(2)\}$, $\{\Gamma_0(3), \Gamma^0(3)\}$, $\{\Gamma_0(4), \Gamma^0(4)\}$\\ \hline 
    \end{tabular}
    \caption{Possible congruence subgroups of $\{SL(2,\mathbb{Z})_1, SL(2,\mathbb{Z})_2\}$ 
    except for the trivial pattern $a^1=a^2$. 
    Those are constrained by the tadpole cancellation condition~(\ref{eq:tadT6Z2}) 
    and quantization condition of flux quanta $a^1,a^2 \in 2\mathbb{Z}$. Here, we restrict ourselves to the case with $a^2/a^1> 1$, but it is also possible to consider the case with the replacement $\Gamma^{(2)}$ of $\Gamma^{(1)}$.}
    \label{tab:T6Z2}
\end{table}

In this way, the low-energy effective action has discrete modular symmetries 
below the mass scale of heavy modulus. 
The remaining modular group in the low-energy effective action is severely 
constrained by the quantization of 
fluxes and tadpole cancellation condition. 
We have focused on the specific moduli stabilization, but we 
expect that allowed congruence subgroups are 
restricted in a general choice of flux quanta on $T^6/\mathbb{Z}_2$ 
due to the quantization of three-form fluxes (\ref{eq:T6Z2quant}). 
In the following, we show other examples as the generalization of our setup.

\subsubsection{Model 2}

So far, our approach has been restricted to the superpotential~(\ref{eq:Wsim}) 
leading to the flat directions in the moduli spaces of the axio-dilaton and 
complex structure moduli.\footnote{For the stabilization of all the 
complex structure moduli and axio-dilaton at supersymmetric and 
supersymmetry-breaking minima, we refer \cite{Honma:2019gzp}.} 
Here the flat direction leading to the discrete modular groups can be achieved in the linear combination of two 
complex structure moduli fields, but it is possible to obtain 
more complicated flat directions of the moduli space in general. 
In this section, we show next non-trivial examples leading to 
the flat direction of the moduli space.

We consider the following superpotential \cite{Hebecker:2017lxm}:
\begin{align}
    W &= (\tau_3 -f\tau)\biggl[ a^0 \tau_1\tau_2 -b_3\biggl],
      \label{eq:Wsim}
\end{align}
by further imposing $a^1=a^2=0$ in Eq. (\ref{eq:Wgeneral}). 
We find the supersymmetric minimum
\begin{align}
    \tau_1 \tau_2 =\frac{b_3}{a^0},\qquad \tau_3=f\tau,
\end{align}
and the flux-induced D3-brane charge is given by
\begin{align}
    N_{\rm flux} = -d_0a^0 +c^3b_3=-2fa^0b^3.
\end{align}

In this case, the three-form fluxes are introduced on the following basis:
\begin{align}
    F_3 &= (a^0dx_1 \wedge dx^2 - b_3 dy^1\wedge dy_2)\wedge dx_3 
    \equiv A_{ij} d\xi_1^i \wedge d\xi_2^j \wedge dx_3,
    \nonumber\\
    H_3 &= (c_3dx_1 \wedge dx^2 + d_0 dy^1\wedge dy_2)\wedge dy_3,    
    \equiv -fA_{ij} d\xi_1^i \wedge d\xi_2^j \wedge dy_3,
\end{align}
with 
\begin{align}
    A_{ij} \equiv 
     \begin{pmatrix}
       -b_3 & 0\\
       0 & a^0
    \end{pmatrix}       
,\qquad 
    \xi_i \equiv 
     \begin{pmatrix}
       y_i\\
       x_i
    \end{pmatrix}       
.
\end{align}
From the modular transformation of the three-form fluxes,
\begin{align}
F_3 &\rightarrow (R_1^{-1} A (R_2^{-1})^T)_{ij} d\xi_1^i \wedge d\xi_2^j \wedge dx_3,
\nonumber\\
H_3 &\rightarrow -f(R_1^{-1} A (R_2^{-1})^T)_{ij} d\xi_1^i \wedge d\xi_2^j \wedge dy_3,
\label{eq:F3H3}
\end{align}
we require $R_1^{-1} A (R_2^{-1})^T=A$ in a way similar to the model 1. 
It can be satisfied under
\begin{align}
    R_2 = A^T (R_1^{-1})^T (A^{-1})^T = 
    \begin{pmatrix}
       t_1 &  s_1 b_3/a^0 \\
       q_1a^0/b_3 & p_1
    \end{pmatrix}
    .
    \label{eq:R2model2}
\end{align}    
It turns out that $q_1$ and $s_1$ are multiples of $b_3$ and $a^0$, respectively. 

We checked that the vacuum condition $\tau_3=\tau$ holds after the modular 
transformation in model 1, where we choose $f=1$. 
It is straightforward to check that 
$\tau_1\tau_2 = b_3/a^0$ also holds after the modular transformations,
\begin{align}
    \tau_1^\prime\tau_2^\prime =R_1(\tau_1)R_2(\tau_2) =\left(\frac{p_1\tau_1+q_1}{s_1\tau_1 +t_1}\right)
    \frac{t_1\tau_2 +s_1b_3/a^0}{(q_1a^0/b_3)\tau_2 +p_1} 
    =\frac{b_3}{a^0}\left(\frac{p_1\tau_1+q_1}{s_1\tau_1 +t_1}\right)
    \frac{t_1/\tau_1+s_1}{q_1/\tau_1 +p_1} =\frac{b_3}{a^0},
    \label{eq:tau1tau2}
\end{align}
where we employ $\tau_1\tau_2 = b_3/a^0$.

As a result, the remaining congruence subgroups are the same as in model 1, as shown in Table~\ref{tab:T6Z2}, when we replace the flux pair $(a^1,a^2)$ in model 1 by the flux pair $(a^0, b^3)$. 
This is because the tadpole cancellation condition as well as the matrix $R_2$ 
has the same structure.

\subsubsection{Model 3}

Here, we discuss other non-trivial examples leading to the congruence 
subgroups in the flat direction of the moduli space. 
Let us consider the following superpotential,
\begin{align}
    W &=  (\tau_3 -f\tau)\biggl[ a^0 \tau_1\tau_2 
    - a^2\tau_1 -b_3\biggl],
\end{align}
by setting $a^1=0$ in Eq.~(\ref{eq:Wgeneral})\footnote{The following discussion is also applicable 
in the case with $a^2=0$ and $a^1\neq 0$.}, indicating that the flux-induced D3-brane charge is given by
\begin{align}
    N_{\rm flux} = -d_0a^0 +c^3b_3=-2fa^0b^3.
\end{align}
The supersymmetric minimum $\partial_\tau W = \partial_{\tau_i} W=W=0$ with $i=1,2,3$ is realized at
\begin{align}
     &\tau_3 = f\tau,\qquad 
     a^0\tau_1\tau_2 -a^2\tau_1 -b_3= 0.
     \label{eq:vacmodel3}
\end{align}
In a way similar to the analyses in the previous subsections, 
we discuss the congruence subgroups on the above flat directions of the moduli space. 
Since the $\tau_3=f\tau$ direction is the same with models 1 and 2, we focus on 
the other flat direction. 
The three-form fluxes are introduced as follows:
\begin{align}
    F_3 &= (a^0dx_1 \wedge dx^2 +a^2dx_1 \wedge dy^2- b_3 dy^1\wedge dy_2)\wedge dx_3 
    \equiv A_{ij} d\xi_1^i \wedge d\xi_2^j \wedge dx_3,
    \nonumber\\
    H_3 &= (c_3dx_1 \wedge dx^2 -d_1dx_1 \wedge dy^2+ d_0 dy^1\wedge dy_2)\wedge dy_3,    
    \equiv -fA_{ij} d\xi_1^i \wedge d\xi_2^j \wedge dy_3,
\end{align}
with 
\begin{align}
    A_{ij} \equiv 
     \begin{pmatrix}
       -b_3 & 0\\
       a^2 & a^0
    \end{pmatrix}       
,\qquad 
    \xi_i \equiv 
     \begin{pmatrix}
       y_i\\
       x_i
    \end{pmatrix}       
.
\end{align}
We require $R_1^{-1} A (R_2^{-1})^T=A$  to keep the invariance of the three-form fluxes under the modular transformations on $SL(2,\mathbb{Z)}_1 \times SL(2,\mathbb{Z})_2$.
We find that it can be achieved when
\begin{align}
    R_2 = A^T (R_1^{-1})^T (A^{-1})^T = 
    \begin{pmatrix}
       1 &   \frac{a^2}{b_3}\frac{b_3}{a^0} \\
       0 & 1
    \end{pmatrix}
    \begin{pmatrix}
       t_1 &  s_1 b_3/a^0 \\
       q_1a^0/b_3 & p_1
    \end{pmatrix}
    \begin{pmatrix}
       1 &   -\frac{a^2}{b_3}\frac{b_3}{a^0} \\
       0 & 1
    \end{pmatrix}
    .
    \label{eq:R2model3}
\end{align}    
In this way, it is the generalization of (\ref{eq:R2model2}). 
When $a^2/b_3 \in \mathbb{Z}$ and $b_3/a^0 \in \mathbb{Z}$, Eq.~(\ref{eq:R2model3}) is the element of $\Gamma^0(b_3/a^0)$; otherwise, it is difficult to obtain the 
congruence subgroups. 
Then, $q_1$ and $s_1$ are required to be multiples of $b_3$ and $a^0$, respectively. 
Since the tadpole cancellation condition as well as the structure of the matrix 
$R_2$ is the same with models 1 and 2, the remaining modular group in 
the effective action is described by Table~\ref{tab:T6Z2}. 
Note that it is straightforward to check the modular invariance of the vacuum condition (\ref{eq:vacmodel3}) 
by employing Eqs.~(\ref{eq:R2model3}) and (\ref{eq:vacmodel3}).

So far, we have discussed the congruence subgroups as a special case of Eq.~(\ref{eq:Wgeneral}) by setting certain fluxes to be 0. 
As a result, flat direction in Eq.~(\ref{eq:vacuumGeneral}) 
has the congruence subgroups enumerated in Eqs.~(\ref{eq:Congruence1})-(\ref{eq:Congruence3}). 
The important point to realize the (discrete) modular groups in the effective 
action is that the three-form fluxes are expanded on the basis $\xi$ which manifests the 
modular invariance of $F_3$ and $H_3$. If the basis of $F_3$ and $H_3$ explicitly depends on the coordinates $x$ and $y$ 
($x_3, y_3$ in the previous setup (\ref{eq:F3H3})), the modular transformations change the $G_3$ itself 
and $SL(2,\mathbb{Z})_\tau$ transformation is required to compensate the transformation of $G_3$ 
as demonstrated in $\tau_3=\tau$ direction (\ref{eq:tautau3}). 
Hence, we expect that the flat directions possessing the modular symmetries in the superpotential (\ref{eq:W}) 
are typically characterized by the two moduli fields among $(\tau_1, \tau_2, \tau_3, S)$, 
and our discussed superpotential is a representative one. 
This argument will hold not only $T^6/\mathbb{Z}_2$, but also more general toroidal orientifolds as well. 

Let us briefly comment on other toroidal orientifolds. 
Compared with the $T^6/\mathbb{Z}_2$ and $T^6/(\mathbb{Z}_2\times \mathbb{Z}_2^\prime)$ orientifolds, 
the untwisted complex structure moduli are fixed at discrete values or 
described by the single modulus on other toroidal orbifolds preserving the supersymmetry such as 
$T^6/(\mathbb{Z}_2\times \mathbb{Z}_3)=T^6/\mathbb{Z}_{6-II}$, $T^6/(\mathbb{Z}_2\times \mathbb{Z}_6)$, 
$T^6/\mathbb{Z}_4$, $T^6/\mathbb{Z}_{8-II}$, and 
$T^6/\mathbb{Z}_{12-II}$~\cite{Ibanez:1987pj,Font:1988mk,Katsuki:1989bf,Kobayashi:1991rp}. 
The presence of modular symmetry on other toroidal orientifolds is a restricted class of $T^6/\mathbb{Z}_2$ 
orientifold, taking into account the tadpole cancellation conditions as well as the quantization condition of fluxes.

\section{Modular symmetry on $T^6/(\mathbb{Z}_2\times \mathbb{Z}_2^\prime)$ orientifolds with magnetized D-branes}
\label{sec:3}
So far, we have not considered the matter sector. 
In this section, we introduce the magnetized D-branes on 
$T^6/(\mathbb{Z}_2\times \mathbb{Z}_2^\prime)$ orientifolds 
rather than $T^6/\mathbb{Z}_2$. 
Similar to the factorizable 6-torus $T^6/\mathbb{Z}_2$ discussed in the previous section, 
the three-forms can be expanded on the same basis in (\ref{eq:basis}) which 
are invariant under the $\mathbb{Z}_2\times \mathbb{Z}_2^\prime$ orbifoldings. 
In this way, we can apply the moduli stabilization scheme of Section~\ref{sec:2}, 
in particular model 1 in Sec.~\ref{sec:model1}, 
to $T^6/(\mathbb{Z}_2\times \mathbb{Z}_2^\prime)$ background. 
It indicates that the congruence subgroups also appear in the 4D effective action, 
although the tadpole cancellation condition and the quantization 
condition of fluxes are modified due to the 
orientifold contributions and inclusion of the D-branes with magnetic fluxes.

\subsection{$T^6/(\mathbb{Z}_2\times \mathbb{Z}_2^\prime)$ orientifold models}
In this section, we review the $T^6/(\mathbb{Z}_2\times \mathbb{Z}_2^\prime)$ 
orientifold models with or without discrete torsion, where the semi-realistic phenomenological models are 
previously found in D-branes with magnetic fluxes~\cite{Blumenhagen:2003vr,Cascales:2003zp,Marchesano:2004xz,Cvetic:2005bn}.

On this background, two $\mathbb{Z}_2$ symmetries act on the $T^6$ coordinates as
\begin{align}
    \theta\,:\,(z_1, z_2, z_3) \rightarrow (-z_1, -z_2, z_3),\qquad
    \theta^\prime\,:\,(z_1, z_2, z_3) \rightarrow (z_1, -z_2, -z_3),    
\end{align}
and the orientifold projection is characterized by the world-sheet parity 
projection $\Omega$ and 
\begin{align}
    {\cal R}\,:\,(z_1, z_2, z_3) \rightarrow (-z_1, -z_2, -z_3).        
\end{align}
Under those actions, there exist 64 O$3$-planes located at a fixed point of ${\cal R}$ and 
4 O$7_1$-, 4 O$7_2$-, 4 O$7_3$-planes, located at the fixed locus of ${\cal R}\theta^\prime$, ${\cal R}\theta\theta^\prime$ and ${\cal R}\theta$, respectively. 

In addition, it is possible to consider $N_a$ stacks of magnetized D$(3+2n)$-branes wrapping $2n$-cycles on 
$T^6/(\mathbb{Z}_2\times \mathbb{Z}_2^\prime)$, where $U(1)_a$ magnetic fluxes $F_a$ 
are quantized on $T_i^2$, 
\begin{align}
    \frac{m_a^i}{2\pi} \int_{T_i^2} F_a^i = n_a^i.
\end{align}
Here the integer $m_a^i$ denotes the wrapping number of $N_a$ D$(3+2n)$-branes around $T_i^2$ and 
$n_a^i$ are quantized fluxes. 
Note that D3-, D5-, D7-, D9-branes 
can consider 0, 1, 2, and, 3 non-vanishing $m_a^i$ fluxes, respectively. 
Under the orientifold projection, the wrapping number $m_a^i$ transforms as 
$\Omega {\cal R}\,:\,m_a^i\rightarrow -m_a^i$. 
Such gauge fluxes not only break the original gauge symmetry 
of D$(3+2n)$-branes, but also induce the chiral zero-modes 
at the intersection of two stacks of D-branes, counted by
\begin{align}
    I_{ab} = \Pi_{i=1}^3 (n_a^i m_b^i -n_b^i m_a^i),
\end{align}
where the labels $a$ and $b$ represent two stacks of D-branes. 

Recall that the above magnetic fluxes also carry the Ramond-Ramond (RR) 
charges of lower dimensional D-branes through the Chern-Simons coupling. 
Thanks to the orientifold projection, RR tadpoles of D5- and D9-branes 
are cancelled by their orientifold images. 
It is thus required to take into account only the D3- and D7-brane 
charges~\cite{Marchesano:2004xz}:
\begin{align}
    {\rm D}3\;&:\; \sum_a N_a n_a^1n_a^2n_a^3 +\frac{1}{2}N_{\rm flux} =16,
    \nonumber\\
    {\rm D}7_1\;&:\; \sum_a N_a n_a^1m_a^2m_a^3 =-16,
    \nonumber\\
    {\rm D}7_2\;&:\; \sum_a N_a n_a^2m_a^1m_a^3 =-16,
    \nonumber\\
    {\rm D}7_3\;&:\; \sum_a N_a n_a^3m_a^1m_a^2 =-16,
    \label{eq:TadT6Z2Z2}
\end{align}
where O3- and O$7_i$-planes have -1/2 units of D3-brane charge and 
-8 units of D$7_i$-brane charge, respectively.

Our interest is to reveal the allowable values of flux quanta $N_{\rm flux}$, 
determining the breaking of the modular symmetry as discussed in 
the previous section. 
On $T^6/(\mathbb{Z}_2\times \mathbb{Z}_2^\prime)$ background, one can introduce 
the discrete torsion in $\mathbb{Z}_2$ twisted sector~\cite{Vafa:1986wx,Vafa:1994rv,Douglas:1998xa}, 
which changes a part of the orientifold charges and the quantization condition of fluxes. 
The quantization of the three-form fluxes on $\mathbb{Z}_2 \times \mathbb{Z}_2^\prime$ with discrete torsion\footnote{We use the 
conventions that the hodge numbers of $T^6/(\mathbb{Z}_2\times \mathbb{Z}_2^\prime)$ are 
$(h_{1,1}, h_{2,1})=(3, 51)$ with discrete torsion and $(h_{1,1}, h_{2,1})=(51, 3)$ without discrete torsion, 
respectively. Furthermore, we focus on the untwisted sector on $T^6/(\mathbb{Z}_2\times \mathbb{Z}_2^\prime)$ with discrete torsion.}, 
the three-form fluxes $F_3$ and $H_3$ are quantized in multiples of 8, due to the 
$\mathbb{Z}_2\times \mathbb{Z}_2$ orbifold and orientifold 
projections. 
The reason is that the volume of a three-cycle on $T^6$ is divided by the 
corresponding cycle on $T^6/(\mathbb{Z}_2 \times \mathbb{Z}_2^\prime)$~\cite{Blumenhagen:2003vr} 
and the orientifold $\mathbb{Z}_2$ projection further act on $T^6$~\cite{Frey:2002hf}. 
On the other hand, in the case without discrete torsion, 
$F_3$ and $H_3$ are quantized in multiples of 4. (For more details, see, \cite{Blumenhagen:2003vr}, 
in which the convention of discrete torsion is opposite to ours.)

\subsection{Congruence subgroups}
As mentioned before, the three-form basis on $T^6/(\mathbb{Z}_2 \times \mathbb{Z}_2^\prime)$ 
can be expanded in terms of Eq.~(\ref{eq:basis}). 
It indicates that we can apply the moduli stabilization mechanism of Section~\ref{sec:2} 
into this $T^6/(\mathbb{Z}_2 \times \mathbb{Z}_2^\prime)$ orientifold 
background, taking into account the tadpole cancellation 
conditions~(\ref{eq:TadT6Z2Z2}). As classified later, the three-form fluxes 
lead to the congruence subgroups on $T_1^2$ and/or $T_2^2$. After integrating out 
the heavy modulus, the low-energy effective action is invariant under the remaining 
discrete subgroups of $SL(2,\mathbb{Z})$. 
In the following, we enumerate the possible congruence subgroups on $T^6/(\mathbb{Z}_2 \times \mathbb{Z}_2^\prime)$ with or without discrete torsion.

\begin{itemize}
\item
We first discuss $T^6/(\mathbb{Z}_2 \times \mathbb{Z}_2^\prime)$ 
with discrete torsion. Given that $F_3$ and $H_3$ are quantized 
in multiples of 8, namely $a^1, a^2 \in 8 \mathbb{Z}$, we obtain the class of 
congruence subgroups of $SL(2, \mathbb{Z})_1 
\times SL(2, \mathbb{Z})_2$ modular group on $T_1^2 \times T_2^2$, 
displayed in Table~\ref{tab:T6Z2Z21}. Here, we employ the model 1 
in Sec.~\ref{sec:model1}. 
When both the quark and lepton sectors arise in D-branes wrapping $T_1^2$ ($T_2^2$), their flavor symmetries are governed by the same $\Gamma^{(1)}$ ($\Gamma^{(2)}$). 
On the other hand, when their flavor symmetries are originated from different tori, 
for instance, the quark sector on $T_1^2$ and the lepton sector on $T_2^2$, 
quarks and leptons have the different flavor symmetries 
$\Gamma^{(1)}$ and $\Gamma^{(2)}$ as shown in Table~\ref{tab:T6Z2Z21}, respectively. 
Thus, our results would be interesting for bottom-up model building studied in Refs.~\cite{Kobayashi:2018wkl,deMedeirosVarzielas:2019cyj}.

\begin{table}[htb]
    \centering
    \begin{tabular}{|c|c|} \hline
   $(f, a^1, a^2)$    & $\{\Gamma^{(1)}, \Gamma^{(2)}\}$\\ \hline \hline
   $(1, \mp 8, \pm 16)$    & $\{\Gamma^0(2), \Gamma_0(2)\}$\\ \hline 
   $(1, \mp 16, \pm 8)$    &  $\{\Gamma_0(2), \Gamma^0(2)\}$\\ \hline  
    \end{tabular}
    \caption{Possible congruence subgroups of $\{SL(2,\mathbb{Z})_1, SL(2,\mathbb{Z})_2\}$ 
    except for the trivial $a^1=a^2$ case, up to $N_{\rm flux}=-2fa^1a^2=64\times 4$. 
    Since $N_{\rm flux}$ is multiples of 64 due to the 
    flux quantization of $F_3$ and $H_3$, only $N_{\rm flux}=64\times 4$ gives rise to 
    the non-trivial congruence subgroup $\Gamma^0(2)$ and $\Gamma_0(2)$.}
    \label{tab:T6Z2Z21}
\end{table}

\item
Next, we focus on the case without discrete torsion, in which 
flux quanta $a^1$ and $a^2$ are quantized in multiples of 4, 
namely $a^1, a^2 \in 4 \mathbb{Z}$. 
In a similar way, the flavor symmetries on D-branes wrapping 
$T_1^2$ and/or $T_2^2$ are governed by the congruence subgroups displayed in Table~\ref{tab:T6Z2Z22}. 
We have more varieties than congruence groups in Tables \ref{tab:T6Z2} and \ref{tab:T6Z2Z21}.
We can consider the same or different flavor symmetries of quark and lepton sectors, 
depending on the D-brane configurations.

\begin{table}[htb]
    \centering
    \begin{tabular}{|c|} \hline
  $\{\Gamma^{(1)}, \Gamma^{(2)}\}$\\ \hline \hline
   $\{\Gamma_0(2), \Gamma^0(2)\}$, $\{\Gamma_0(3), \Gamma^0(3)\}$, 
   $\{\Gamma_0(4), \Gamma^0(4)\}$, $\{\Gamma_0(5), \Gamma^0(5)\}$, 
   $\{\Gamma_0(6), \Gamma^0(6)\}$, $\{\Gamma(6), \Gamma(6)\}$\\ \hline 
    \end{tabular}
    \caption{Possible congruence subgroups of $\{SL(2,\mathbb{Z})_1, SL(2,\mathbb{Z})_2\}$ 
    except for the trivial $a^1=a^2$ case, up to $N_{\rm flux}=-2fa^1a^2= 16\times 12$, where $N_{\rm flux}$ is multiples of 16 due to the 
    flux quantization of $F_3$ and $H_3$. Here, we restrict ourselves to 
    the $|a^2|>|a^1|$ case, but it is also possible to consider the $|a^1|>|a^2|$ case.}
    \label{tab:T6Z2Z22}
\end{table}
\end{itemize}

As a result, on $T^6/(\mathbb{Z}_2 \times \mathbb{Z}_2^\prime)$ orientifold 
background, the congruence subgroups of the modular group in the low-energy 
effective action is 
severely constrained by the flux quantization and tadpole cancellation 
conditions.　

\subsection{Concrete models}

\subsubsection{Model with discrete torsion}
\label{sec:3_3_1}
In this section, we search for a concrete three-generation model including the discrete 
modular symmetry in a specific 
D-brane configuration on the $T^6/(\mathbb{Z}_2\times \mathbb{Z}_2^\prime)$ 
orientifold background with discrete torsion, following \cite{Marchesano:2004xz}.

Massless spectrum on the stack $a$ of magnetized D-branes 
has the representation of $U(N_a/2)$ not $U(N_a)$ due to 
the orbifold projection, meaning that $N_a$ must be even. 
There exist ${\cal N}=1$ $U(N_a/2)$ vector multiplet and three 
chiral multiplets (open string moduli), which we refer as a $(aa)$ 
sector. 
By taking into account the chiral spectrum between two stacks 
$a$ and $b$ of D-branes, the massless spectrum for magnetized D-branes 
is summarized as follows:
\begin{align}
    aa\,{\rm sector}:&\, 
    \left\{
    \begin{array}{c}
    U(N_a/2)\,{\rm vector~multiplet}\\
    3{\rm Adj.~chiral~multiplets}   
    \end{array}
\right.
,
\\ \nonumber
    ab + ba\,{\rm sector}:&\, I_{ab}\,(\fund_a, \antifund_b)\,{\rm chiral~multiplets},
    \\ \nonumber
    ab^\prime + b^\prime a\,{\rm sector}:&\, I_{ab^\prime}\,(\fund_a, \fund_b)\,{\rm chiral~multiplets},
    \\ \nonumber
    aa^\prime + a^\prime a\,{\rm sector}:&\,
    \left\{
    \begin{array}{c}
\frac{1}{2}\left(I_{aa^\prime}-4I_{a,O}\right)\,\Ysymm\,{\rm chiral~multiplets}
    \\ \nonumber
    \frac{1}{2}\left(I_{aa^\prime}-4I_{a,O}\right)\,\Yasymm\,{\rm chiral~multiplets}
    \end{array}
\right.
,
\end{align}
where $a^\prime$ denotes the $\Omega {\cal R}$ image of the stack $a$ of D-brane and 
$I_{a,O}$ represents for the intersection product between the homology class of D-branes 
and and orientifold plane. 
In addition, it is also possible to consider D-branes whose homology class is invariant 
under the orientifold projection for the configuration of D3 and D$7_i$-branes without magnetic fluxes. 
In contrast to the case without fixed by the orientifold action, the 
Chan-Paton gauge group for a stack of $2N_a$ D-branes is described by $USp(N_a)$ 
gauge group. The massless spectrum is summarized as follows:
\begin{align}
    aa\,{\rm sector}:&\,
        \left\{
    \begin{array}{c}USp(N_a)\,{\rm vector~multiplet}\\
    3\, \Yasymm \,{\rm chiral~multiplets}
    \end{array}
    \right.
    ,
    \\ \nonumber
    ab + ba\,{\rm sector}:&\,\, I_{ab}\,(\fund_a, \antifund_b)\,{\rm chiral~multiplets}.
\end{align}

Let us engineer the brane configurations leading to the semi-realistic model accommodating 
the Standard Model. 
The visible sector we consider consists of stacks of magnetized D$7_a$ with 
gauge group $U(4)_a$ and two stacks of unmagnetized D$7_b$ and D$7_c$-branes with $SU(2)_L\times SU(2)_R$ as shown in Table~\ref{tab:Magneticflux}.
\begin{table}[]
    \centering
    \begin{tabular}{|c|c|c|c|c|} \hline
        $N_\alpha$ & {\rm Gauge~group} & ($n_\alpha^1, m_\alpha^1$) & ($n_\alpha^2, m_\alpha^2$) & ($n_\alpha^3, m_\alpha^3$)\\ \hline 
        $N_a=8$ & $U(4)_a$ & (1,0) & ($g, 1$) & ($g, -1$)\\ \hline
         $N_b=2$ & $SU(2)_L$ & (0,1) & ($1, 0$) & ($0, -1$)\\ \hline
        $N_c=2$ & $SU(2)_R$ & (0,1) & ($0, -1$) & ($1,0$)\\ \hline\hline
        $N_{h_1}=2$ & $U(1)_{h_1}$ & (-2,-1) & (3, 1) & (4,1)\\ \hline
        $N_{h_2}=2$ & $U(1)_{h_2}$ & (-2,-1) & (4, 1) & (3,1)\\ \hline
        $8N_f$ & $USp(8N_f)$ & (1,0) & (1,0) & (1,0)\\ \hline
    \end{tabular}
    \caption{D-brane configurations leading to Pati-Salam model where the magnitude of magnetic fluxes $g$ corresponds to the generations chiral multiplets in the visible sector. Here, we use the identification $USp(2)\simeq SU(2)$.}
    \label{tab:Magneticflux}
\end{table}
To cancel RR tadpoles, we also introduce two stacks of magnetized D9-branes with $U(1)_{h_1}\times U(1)_{h_2}$ and 
the 8$N_f$ D3-branes with $USp(8N_f)$ which are  located at the top of an orientifold singularity.

By taking into account the Green-Schwarz mechanism, cancelling the gauge and gravitational anomalies, 
two $U(1)$ gauge bosons become massive at a compactification scale.
Hence, remaining gauge symmetry consists of 
the Pati-Salam and hidden sectors:
\begin{align}
    SU(4) \times SU(2)_L \times SU(2)_R \times U(1)^\prime \times USp(8N_f)
\end{align}
with $U(1)^\prime = U(1)_a -2g(U(1)_{h_1}-U(1)_{h_2})$. 
The latter two are the hidden sector. 
After all, cancellation condition of D3-brane charge~(\ref{eq:TadT6Z2Z2}) requires that 
\begin{align}
    g^2 + N_f + \frac{N_{\rm flux}}{16} =14.
\end{align}
When we focus on the three-generation model i.e. $g=3$, we have two options to satisfy the above, 
namely $\{N_{\rm flux}=0, N_f=5\}$ or $\{N_{\rm flux}=64, N_f=1 \}$. 
Vanishing three-form fluxes $N_{\rm flux}=0$ correspond to the T-dual type IIA intersecting D-brane model~\cite{Cremades:2003qj}. 

As demonstrated in the previous subsection, 
it is possible to break the modular group to its subgroup 
on the non-trivial three-form background. 
However, in the $N_{\rm flux}=64$ case, we cannot apply our moduli stabilization 
scheme to this case due to $N_{\rm flux}=-2f a^1a^2 \in 128\mathbb{Z}$ with $a^1,a^2 \in 8\mathbb{Z}$ and $f\in \mathbb{Z}$. 
In the next section, we explore the possibility to realize the congruence subgroup 
in the toroidal orientifold model without discrete torsion.

\subsubsection{Model without discrete torsion}
\label{sec:3_3_2}
The purpose of this section is to explore the existence of the congruence subgroups on $T^6/(\mathbb{Z}_2\times \mathbb{Z}_2^\prime)$ orientifold 
models without discrete torsion. 
For our purpose, we consider the specific 
semi-realistic models, developed in Ref.~\cite{Blumenhagen:2005tn}. 

Along the line of Ref.~\cite{Blumenhagen:2005tn}, we choose 
exotic 64 O$3^{(+,+)}$-planes with positive charge and tension located at the fixed 
points of orbifold and 
orientifold actions. The 4 O$7_i$-planes are located 
at the fixed locus of ${\cal R}\theta^\prime$, ${\cal R}\theta\theta^\prime$ and ${\cal R}\theta$, in 
the same way as the model with discrete torsion. 
Such exotic O3-plane contributions change the tadpole cancellation 
condition of D3-brane charge:
\begin{align}
\sum_a N_a n_a^1n_a^2n_a^3 +\frac{1}{2}N_{\rm flux} =-16,    
\end{align}
whereas the cancellation condition of D$7_i$-brane charges is the same with 
Eq.~(\ref{eq:TadT6Z2Z2}), taking 
into account the extra K-theory constraints. 
To demonstrate the semi-realistic models, 
we consider the magnetic fluxes and wrapping numbers of D$(3+2n)$-branes in 
Table~\ref{tab:mflux_wotorsion}, where the visible sector is constructed 
on fractional D7-branes and D9-${\rm \bar{D}}$9 pairs. 
In particular, the visible sector consists of  four-generation ${\cal N}=1$ supersymmetric Pati-Salam-like model with 
$SU(4)_C\times SU(2)_L\times SU(2)_R$ gauge 
groups, where some of the $U(1)$ gauge bosons become massive through the Green-Schwarz mechanism. 
For more details on the model, see Ref.~\cite{Blumenhagen:2005tn}.

\begin{table}[h]
    \centering
    \begin{tabular}{|c|c|c|c|c|c|} \hline
        & $N_\alpha$ & {\rm Gauge~group} &   ($n_\alpha^1, m_\alpha^1$) & ($n_\alpha^2, m_\alpha^2$) & ($n_\alpha^3, m_\alpha^3$)\\ \hline 
        Fractional & $N_{a_1}=4$ & $U(4)_C$ & (1,0) & ($0, 1$) & ($0, -1$)\\ \cline{2-6}
        D-branes   & $N_{a_2}=2$ & $U(2)_L$ & (1,0) & ($2, 1$) & ($4, -1$)\\ \cline{2-6}
        (visible sector) & $N_{a_3}=2$ & $U(2)_R$ & (-3,2) & ($-2, 1$) & ($-4,1$)\\ \hline\hline
        Bulk & $N_{b}=4$ & $U(2)^2$ & (1,0) & ($0, 1$) & ($0, -1$)\\ \cline{2-6}
        D-branes   & $N_{c}=8$ & $U(4)^2$ & (0,1) & ($1, 0$) & ($0, -1$)\\ \cline{2-6}
        & $N_{d}=8N_f$ & $USp(4N_f)^4$ & (1,0) & (1,0) & (1,0)\\ \hline
    \end{tabular}
    \caption{D-brane configurations leading to Pati-Salam model.} 
    \label{tab:mflux_wotorsion}
\end{table}

Let us classify the congruence subgroups on this setup by introducing the three-form fluxes as in 
Section~\ref{sec:3_3_1}. 
From the wrapping numbers in Table~\ref{tab:mflux_wotorsion}, the tadpole cancellation condition of D3-brane charges 
reads
\begin{align}
    N_f +\frac{N_{\rm flux}}{16} =2.
\end{align}
Here, $N_{\rm flux}$ is quantized in multiples of 16, due to the quantization condition of $F_3$ and $H_3$. 
We have three options to satisfy the above, 
namely $\{N_{\rm flux}=0, N_f=2\}$, $\{N_{\rm flux}=16, N_f=1 \}$ and $\{N_{\rm flux}=32, N_f=0 \}$. 
The stabilization mechanism in Section~\ref{sec:model1} leads to $N_{\rm flux}=-2f a^1a^2$ with $a^1,a^2 \in 4\mathbb{Z}$, 
indicating that only $N_{\rm flux}=32$ is consistent with the tadpole cancellation condition. 
In that case, we have the unique possibility $|a^1|=|a^2|=4$ and $f=1$. 
Thus, the diagonal parts of $SL(2,\mathbb{Z})_1 \times SL(2,\mathbb{Z})_2$ and $SL(2,\mathbb{Z})_3 \times SL(2,\mathbb{Z})_\tau$ remain 
in the effective action.

In this specific Pati-Salam model having a quark-lepton unification, flavor structures of quarks and leptons are determined by the magnetic fluxes inserted on $T_2^2\times T_3^2$, indicating that both quarks and leptons 
transform under the same modular symmetry of $T_2^2\times T_3^2$. 
As a result, their flavor symmetries on $T_2^2$ could be determined by $SL(2,\mathbb{Z})$. 

Although we have focused on the specific D-brane configurations yielding the 
modular group in the effective action, it is interesting to explore D-brane models 
accommodating not only the three-generation models, but also the finite modular 
symmetry shown in Table~\ref{tab:T6Z2Z22}.

\section{Conclusions and Discussions}
\label{sec:conclusion}

In this paper, we have classified possible congruence subgroups of the modular 
group in the effective action of Type IIB string theory on toroidal 
orientifolds with three-form fluxes. 
The discrete modular symmetry arises in the flat direction of 
complex structure moduli whose moduli space has a congruence 
subgroup rather than $SL(2,\mathbb{Z})$. 
The realization of discrete modular group in the effective action 
has been achieved by an existence of three-form fluxes inserted on 3-cycles of $T^6$. 
We argued that such a discrete modular group plays an important role of 
not only enlarging the axionic field range discussed in the context of 
Swampland conjecture~\cite{Hebecker:2017lxm}, but also the flavor symmetry of quarks and leptons.

Indeed, when magnetized D-branes wrap a certain cycle of tori, 
we could identify the remaining discrete modular symmetry 
with the flavor symmetry of quarks and/or leptons. 
We discussed the possible congruence subgroups on 
$T^6/(\mathbb{Z}_2\times \mathbb{Z}_2^\prime)$ orientifold with and 
without discrete torsion, incorporating the standard model sector. 
It turned out that the possible class of congruence subgroups are sensitive to 
the quantization of fluxes and the tadpole cancellation conditions. 
We expect that our analysis would be applicable to more broad class of 
D-brane model building on toroidal orientifolds. 
It is interesting to clarify the congruence subgroups 
of modular symmetry on other toroidal orientifolds (discussed in e.g., Ref.~\cite{Lust:2005dy}) 
as well as non-factorizable tori incorporating $SL(4, \mathbb{Z})$ 
or $SL(6,\mathbb{Z})$ using the method in this paper.\footnote{Non-factorizable magnetic fluxes 
\cite{Antoniadis:2009bg,Abe:2014nla} would also be interesting.} 
In this paper, we focused on Type IIB flux compactifications, but our discussion is 
also applicable to the Type IIA and Heterotic string flux compactifications, taking into account the similar 
flux superpotential~(\ref{eq:Wgeneral}) and corresponding tadpole cancellation conditions. 
Furthermore, it would be also connected with 
the supergravity models having the 
no-scale property~\cite{Cremmer:1983bf}, 
where the moduli spaces are described by coset spaces such as $SU(p,q)/(U(1)\times SU(p)\times SU(q))$ and $SO(2,2+p)/(SO(2)\times SO(2+p))$. 
Note that the no-scale property holds for 
not only the complex structure moduli in the large complex-structure limit, but also the 
K\"ahler moduli in the large-volume limit. 

We have studied the geometrical symmetry, which is the full symmetry in 
4D low-energy effective supergravity including closed string modes.
On the other hand, matter zero-modes corresponding to quarks and leptons are originated from open string modes.
The number of matter zero-modes is finite, and they transform under the congruence subgroups of the modular symmetry, 
which we have studied.
Such a finite number of zero-modes represent  the remaining modular symmetries, which can be 
finite subgroups.
For example, in Ref.~\cite{Kobayashi:2018rad} it was shown that 
modular symmetries of matter zero-modes are discrete and finite among $SL(2, \mathbb{Z})$ 
 in compactification with magnetic flux.
They depend on the magnitude of magnetic fluxes.
It is very important to study such modular symmetries of zero-modes starting with 
our models, where the full modular symmetry is congruence subgroups.
Depending on the magnitude of magnetic fluxes, we may obtain 
various finite modular subgroups different from those in Ref.~\cite{Kobayashi:2018rad}.
That would provide new 
insights into phenomenological models with discrete flavor 
symmetry. 
We will study these issues elsewhere.

\subsection*{Acknowledgements}

T. K. was supported in part by MEXT KAKENHI Grant Number JP19H04605. 
H. O. was supported in part by JSPS KAKENHI Grant Numbers JP19J00664 and JP20K14477.
\appendix

\section{Congruence subgroups}\label{app}

In this Appendix, we show the conventions of the modular groups we employed. 
The principal congruence subgroup of level $N\in \mathbb{Z}^+$ is
\begin{align}
    \Gamma(N) =
    \left\{
    \begin{bmatrix}
     a & b\\
     c & d
    \end{bmatrix}
    \in SL(2,\mathbb{Z})
    \biggl|
    \begin{bmatrix}
     a & b\\
     c & d
    \end{bmatrix}
    \equiv
    \begin{bmatrix}
     1 ({\rm mod}\,N) & 0({\rm mod}\,N)\\
     0({\rm mod}\,N) & 1({\rm mod}\,N)
    \end{bmatrix}
    \right\}
\end{align}

Definition : A subgroup $\Gamma$ of $SL(2, \mathbb{Z})$ is a {\bf congruence subgroup} if $\Gamma(N) \subset \Gamma$ for some $N\in \mathbb{Z}^+$, in which case $\Gamma$ is a congruence subgroup of level $N$.

The congruence subgroups of Hecke type are
\begin{align}
    \Gamma^0(N) &=
    \left\{
    \begin{bmatrix}
     a & b\\
     c & d
    \end{bmatrix}
    \in SL(2,\mathbb{Z})
    \biggl|
    \begin{bmatrix}
     a & b\\
     c & d
    \end{bmatrix}
    \equiv
    \begin{bmatrix}
     \ast & 0({\rm mod}\,N)\\
     \ast & \ast
    \end{bmatrix}
    \right\}
    ,
    \nonumber\\
    \Gamma_0(N) &=
    \left\{
    \begin{bmatrix}
     a & b\\
     c & d
    \end{bmatrix}
    \in SL(2,\mathbb{Z})
    \biggl|
    \begin{bmatrix}
     a & b\\
     c & d
    \end{bmatrix}
    \equiv
    \begin{bmatrix}
     \ast & \ast\\
     0({\rm mod}\,N) & \ast
    \end{bmatrix}
    \right\},
    \nonumber\\
    \Gamma_1(N) &=
    \left\{
    \begin{bmatrix}
     a & b\\
     c & d
    \end{bmatrix}
    \in SL(2,\mathbb{Z})
    \biggl|
    \begin{bmatrix}
     a & b\\
     c & d
    \end{bmatrix}
    \equiv
    \begin{bmatrix}
     1 ({\rm mod}\,N) & \ast\\
     0({\rm mod}\,N) & 1({\rm mod}\,N)
    \end{bmatrix}
    \right\}
    .
\end{align}

Note that
\begin{align}
    \Gamma(1)  
    =\Gamma_1(1)= \Gamma_0(1)=\Gamma^0(1)=SL(2,\mathbb{Z}).
\end{align}



\begin{thebibliography}{99}
\parskip=-2pt

\bibitem{Peccei:1977hh}
  R.~D.~Peccei and H.~R.~Quinn,
  Phys.\ Rev.\ Lett.\  {\bf 38} (1977) 1440.

\bibitem{Graham:2015cka}
  P.~W.~Graham, D.~E.~Kaplan and S.~Rajendran,
  Phys.\ Rev.\ Lett.\  {\bf 115} (2015) no.22,  221801
  [arXiv:1504.07551 [hep-ph]].

\bibitem{Preskill:1982cy}
  J.~Preskill, M.~B.~Wise and F.~Wilczek,
  Phys.\ Lett.\  {\bf 120B} (1983) 127.

\bibitem{Abbott:1982af}
  L.~F.~Abbott and P.~Sikivie,
  Phys.\ Lett.\  {\bf 120B} (1983) 133.

\bibitem{Dine:1982ah}
  M.~Dine and W.~Fischler,
  Phys.\ Lett.\  {\bf 120B} (1983) 137.  
  


  
\bibitem{Freese:1990rb}
  K.~Freese, J.~A.~Frieman and A.~V.~Olinto,
  Phys.\ Rev.\ Lett.\  {\bf 65} (1990) 3233.
 
\bibitem{Silverstein:2008sg}
  E.~Silverstein and A.~Westphal,
  Phys.\ Rev.\ D {\bf 78} (2008) 106003
  [arXiv:0803.3085 [hep-th]].
  

\bibitem{McAllister:2008hb}
  L.~McAllister, E.~Silverstein and A.~Westphal,
  Phys.\ Rev.\ D {\bf 82} (2010) 046003
  [arXiv:0808.0706 [hep-th]].


\bibitem{Ferrara:1990ei} 
  S.~Ferrara, N.~Magnoli, T.~R.~Taylor and G.~Veneziano,
  Phys.\ Lett.\ B {\bf 245}, 409 (1990).

\bibitem{Cvetic:1991qm} 
  M.~Cvetic, A.~Font, L.~E.~Ibanez, D.~Lust and F.~Quevedo,
  Nucl.\ Phys.\ B {\bf 361}, 194 (1991).

\bibitem{Kobayashi:2016mzg} 
  T.~Kobayashi, D.~Nitta and Y.~Urakawa,
  JCAP {\bf 1608}, 014 (2016)
  [arXiv:1604.02995 [hep-th]].

\bibitem{Hamidi:1986vh} 
  S.~Hamidi and C.~Vafa,
  Nucl.\ Phys.\ B {\bf 279}, 465 (1987);
  L.~J.~Dixon, D.~Friedan, E.~J.~Martinec and S.~H.~Shenker,
  Nucl.\ Phys.\ B {\bf 282}, 13 (1987);
  T.~T.~Burwick, R.~K.~Kaiser and H.~F.~Muller,
  Nucl.\ Phys.\ B {\bf 355}, 689 (1991);
  J.~Erler, D.~Jungnickel, M.~Spalinski and S.~Stieberger,
  Nucl.\ Phys.\ B {\bf 397}, 379 (1993)
  [hep-th/9207049];
  K.~S.~Choi and T.~Kobayashi,
  Nucl.\ Phys.\ B {\bf 797}, 295 (2008)
  [arXiv:0711.4894 [hep-th]].




\bibitem{Cvetic:2003ch} 
  M.~Cvetic and I.~Papadimitriou,
  Phys.\ Rev.\ D {\bf 68}, 046001 (2003)
  [Erratum-ibid.\ D {\bf 70}, 029903 (2004)]
  [hep-th/0303083];
  S.~A.~Abel and A.~W.~Owen,
  Nucl.\ Phys.\ B {\bf 663}, 197 (2003)
  [hep-th/0303124];
%
  D.~Cremades, L.~E.~Ibanez and F.~Marchesano,
  JHEP {\bf 0307}, 038 (2003)
  [hep-th/0302105];
  S.~A.~Abel and A.~W.~Owen,
  Nucl.\ Phys.\ B {\bf 682}, 183 (2004)
  [hep-th/0310257].

\bibitem{Cremades:2004wa}
  D.~Cremades, L.~E.~Ibanez and F.~Marchesano,
  JHEP {\bf 0405} (2004) 079
  [hep-th/0404229];
  H.~Abe, K.~S.~Choi, T.~Kobayashi and H.~Ohki,
  JHEP {\bf 0906}, 080 (2009)
  [arXiv:0903.3800 [hep-th]].






\bibitem{Kobayashi:2017dyu} 
 T.~Kobayashi and S.~Nagamoto,
 Phys.\ Rev.\ D {\bf 96}, no. 9, 096011 (2017)
 [arXiv:1709.09784 [hep-th]].
 
\bibitem{Kobayashi:2018rad} 
 T.~Kobayashi, S.~Nagamoto, S.~Takada, S.~Tamba and T.~H.~Tatsuishi,
 Phys.\ Rev.\ D {\bf 97}, no. 11, 116002 (2018)
 [arXiv:1804.06644 [hep-th]].
 

\bibitem{Kobayashi:2016ovu} 
  T.~Kobayashi, S.~Nagamoto and S.~Uemura,
  PTEP {\bf 2017}, no. 2, 023B02 (2017)
  [arXiv:1608.06129 [hep-th]].

\bibitem{Kobayashi:2018bff}
T.~Kobayashi and S.~Tamba,
Phys.\ Rev.\ D {\bf 99} (2019) no.4, 046001
[arXiv:1811.11384 [hep-th]].

\bibitem{Kariyazono:2019ehj} 
 Y.~Kariyazono, T.~Kobayashi, S.~Takada, S.~Tamba and H.~Uchida,
 Phys.\ Rev.\ D {\bf 100}, no. 4, 045014 (2019)
 [arXiv:1904.07546 [hep-th]].



\bibitem{Baur:2019kwi} 
 A.~Baur, H.~P.~Nilles, A.~Trautner and P.~K.~S.~Vaudrevange,
 Phys.\ Lett.\ B {\bf 795}, 7 (2019)
 [arXiv:1901.03251 [hep-th]]; 
 %
  Nucl.\ Phys.\ B {\bf 947} (2019) 114737
  [arXiv:1908.00805 [hep-th]].
 
\bibitem{Nilles:2020nnc} 
  H.~P.~Nilles, S.~Ramos-Sanchez and P.~K.~S.~Vaudrevange,
  arXiv:2001.01736 [hep-ph].


\bibitem{Feruglio:2017spp} 
  F.~Feruglio,
  arXiv:1706.08749 [hep-ph].

\bibitem{Kobayashi:2018vbk} 
  T.~Kobayashi, K.~Tanaka and T.~H.~Tatsuishi,
  Phys.\ Rev.\ D {\bf 98}, no. 1, 016004 (2018)
  [arXiv:1803.10391 [hep-ph]].

\bibitem{Penedo:2018nmg} 
  J.~T.~Penedo and S.~T.~Petcov,
  Nucl.\ Phys.\ B {\bf 939}, 292 (2019)
  [arXiv:1806.11040 [hep-ph]].

\bibitem{Criado:2018thu} 
  J.~C.~Criado and F.~Feruglio,
  SciPost Phys.\  {\bf 5}, no. 5, 042 (2018)
  [arXiv:1807.01125 [hep-ph]].

\bibitem{Kobayashi:2018scp} 
  T.~Kobayashi, N.~Omoto, Y.~Shimizu, K.~Takagi, M.~Tanimoto and T.~H.~Tatsuishi,
  JHEP {\bf 1811}, 196 (2018)
  [arXiv:1808.03012 [hep-ph]].

\bibitem{Novichkov:2018ovf} 
  P.~P.~Novichkov, J.~T.~Penedo, S.~T.~Petcov and A.~V.~Titov,
  JHEP {\bf 1904}, 005 (2019)
  [arXiv:1811.04933 [hep-ph]].

\bibitem{Novichkov:2018nkm} 
  P.~P.~Novichkov, J.~T.~Penedo, S.~T.~Petcov and A.~V.~Titov,
  JHEP {\bf 1904}, 174 (2019)
  [arXiv:1812.02158 [hep-ph]].

\bibitem{deAnda:2018ecu} 
  F.~J.~de Anda, S.~F.~King and E.~Perdomo,
  arXiv:1812.05620 [hep-ph].

\bibitem{Okada:2018yrn} 
  H.~Okada and M.~Tanimoto,
  Phys.\ Lett.\ B {\bf 791}, 54 (2019)
  [arXiv:1812.09677 [hep-ph]].

\bibitem{Kobayashi:2018wkl} 
  T.~Kobayashi, Y.~Shimizu, K.~Takagi, M.~Tanimoto, T.~H.~Tatsuishi and H.~Uchida,
  Phys.\ Lett.\ B {\bf 794}, 114 (2019)
  [arXiv:1812.11072 [hep-ph]].

\bibitem{Novichkov:2018yse} 
  P.~P.~Novichkov, S.~T.~Petcov and M.~Tanimoto,
  Phys.\ Lett.\ B {\bf 793}, 247 (2019)
  [arXiv:1812.11289 [hep-ph]].

\bibitem{Ding:2019xna} 
  G.~J.~Ding, S.~F.~King and X.~G.~Liu,
  Phys.\ Rev.\ D {\bf 100}, no. 11, 115005 (2019)
  [arXiv:1903.12588 [hep-ph]].

\bibitem{Nomura:2019jxj} 
  T.~Nomura and H.~Okada,
  Phys.\ Lett.\ B {\bf 797}, 134799 (2019)
  [arXiv:1904.03937 [hep-ph]].

\bibitem{Novichkov:2019sqv} 
  P.~P.~Novichkov, J.~T.~Penedo, S.~T.~Petcov and A.~V.~Titov,
  JHEP {\bf 1907}, 165 (2019)
  [arXiv:1905.11970 [hep-ph]].

\bibitem{Okada:2019uoy} 
  H.~Okada and M.~Tanimoto,
  arXiv:1905.13421 [hep-ph].


\bibitem{deMedeirosVarzielas:2019cyj} 
  I.~De Medeiros Varzielas, S.~F.~King and Y.~L.~Zhou,
  arXiv:1906.02208 [hep-ph].

\bibitem{Nomura:2019yft} 
  T.~Nomura and H.~Okada,
  arXiv:1906.03927 [hep-ph].

\bibitem{Kobayashi:2019rzp} 
  T.~Kobayashi, Y.~Shimizu, K.~Takagi, M.~Tanimoto and T.~H.~Tatsuishi,
  arXiv:1906.10341 [hep-ph].

\bibitem{Liu:2019khw} 
  X.~G.~Liu and G.~J.~Ding,
  JHEP {\bf 1908}, 134 (2019)
  [arXiv:1907.01488 [hep-ph]].

\bibitem{Okada:2019xqk} 
  H.~Okada and Y.~Orikasa,
  Phys.\ Rev.\ D {\bf 100}, no. 11, 115037 (2019)
  [arXiv:1907.04716 [hep-ph]].

\bibitem{Kobayashi:2019mna} 
  T.~Kobayashi, Y.~Shimizu, K.~Takagi, M.~Tanimoto and T.~H.~Tatsuishi,
  arXiv:1907.09141 [hep-ph].

\bibitem{Ding:2019zxk} 
  G.~J.~Ding, S.~F.~King and X.~G.~Liu,
  JHEP {\bf 1909}, 074 (2019)
  [arXiv:1907.11714 [hep-ph]].

\bibitem{Okada:2019mjf} 
  H.~Okada and Y.~Orikasa,
  arXiv:1907.13520 [hep-ph].

\bibitem{King:2019vhv} 
  S.~F.~King and Y.~L.~Zhou,
  Phys.\ Rev.\ D {\bf 101}, no. 1, 015001 (2020)
  [arXiv:1908.02770 [hep-ph]].

\bibitem{Nomura:2019lnr} 
  T.~Nomura, H.~Okada and O.~Popov,
  arXiv:1908.07457 [hep-ph].

\bibitem{Okada:2019lzv} 
  H.~Okada and Y.~Orikasa,
  arXiv:1908.08409 [hep-ph].

\bibitem{Criado:2019tzk} 
  J.~C.~Criado, F.~Feruglio, F.~Feruglio and S.~J.~D.~King,
  arXiv:1908.11867 [hep-ph].

\bibitem{Kobayashi:2019xvz} 
  T.~Kobayashi, Y.~Shimizu, K.~Takagi, M.~Tanimoto and T.~H.~Tatsuishi,
  Phys.\ Rev.\ D {\bf 100}, no. 11, 115045 (2019)
  [arXiv:1909.05139 [hep-ph]].

\bibitem{Asaka:2019vev} 
  T.~Asaka, Y.~Heo, T.~H.~Tatsuishi and T.~Yoshida,
  arXiv:1909.06520 [hep-ph].

\bibitem{Gui-JunDing:2019wap} 
  G.~J.~Ding, S.~F.~King, X.~G.~Liu and J.~N.~Lu,
  JHEP {\bf 1912}, 030 (2019)
  [arXiv:1910.03460 [hep-ph]].

\bibitem{Zhang:2019ngf} 
  D.~Zhang,
  arXiv:1910.07869 [hep-ph].

\bibitem{Wang:2019ovr} 
  X.~Wang and S.~Zhou,
  arXiv:1910.09473 [hep-ph].

\bibitem{Kobayashi:2019uyt} 
  T.~Kobayashi, Y.~Shimizu, K.~Takagi, M.~Tanimoto, T.~H.~Tatsuishi and H.~Uchida,
  arXiv:1910.11553 [hep-ph].

\bibitem{Nomura:2019xsb} 
  T.~Nomura, H.~Okada and S.~Patra,
  arXiv:1912.00379 [hep-ph].

\bibitem{Kobayashi:2019gtp} 
  T.~Kobayashi, T.~Nomura and T.~Shimomura,
  arXiv:1912.00637 [hep-ph].

\bibitem{Lu:2019vgm} 
  J.~N.~Lu, X.~G.~Liu and G.~J.~Ding,
  arXiv:1912.07573 [hep-ph].

\bibitem{Wang:2019xbo} 
  X.~Wang,
  arXiv:1912.13284 [hep-ph].







\bibitem{deAdelhartToorop:2011re}
  R.~de Adelhart Toorop, F.~Feruglio and C.~Hagedorn,
  Nucl.\ Phys.\ B {\bf 858} (2012) 437
  [arXiv:1112.1340 [hep-ph]].


\bibitem{Giddings:2001yu} 
  S.~B.~Giddings, S.~Kachru and J.~Polchinski,
  Phys.\ Rev.\ D {\bf 66}, 106006 (2002)
  [hep-th/0105097].







\bibitem{Vafa:2005ui}
  C.~Vafa,
  hep-th/0509212.

\bibitem{Ooguri:2006in}
  H.~Ooguri and C.~Vafa,
  Nucl.\ Phys.\ B {\bf 766} (2007) 21
  [hep-th/0605264].  

\bibitem{Palti:2019pca}
  E.~Palti,
  Fortsch.\ Phys.\  {\bf 67} (2019) no.6,  1900037
  [arXiv:1903.06239 [hep-th]].  



\bibitem{Hebecker:2017lxm}
  A.~Hebecker, P.~Henkenjohann and L.~T.~Witkowski,
  JHEP {\bf 1712} (2017) 033
  [arXiv:1708.06761 [hep-th]].



\bibitem{Blumenhagen:2003vr}
  R.~Blumenhagen, D.~Lust and T.~R.~Taylor,
  Nucl.\ Phys.\ B {\bf 663} (2003) 319
  [hep-th/0303016].

\bibitem{Cascales:2003zp}
  J.~F.~G.~Cascales and A.~M.~Uranga,
  JHEP {\bf 0305} (2003) 011
  [hep-th/0303024].

\bibitem{Marchesano:2004xz}
  F.~Marchesano and G.~Shiu,
  JHEP {\bf 0411} (2004) 041
  [hep-th/0409132].

\bibitem{Cvetic:2005bn}
  M.~Cvetic, T.~Li and T.~Liu,
  Phys.\ Rev.\ D {\bf 71} (2005) 106008
  [hep-th/0501041].



\bibitem{Kachru:2002he}
  S.~Kachru, M.~B.~Schulz and S.~Trivedi,
  JHEP {\bf 0310} (2003) 007
  [hep-th/0201028].



\bibitem{Gukov:1999ya} 
  S.~Gukov, C.~Vafa and E.~Witten,
  Nucl.\ Phys.\ B {\bf 584}, 69 (2000)
  Erratum: [Nucl.\ Phys.\ B {\bf 608}, 477 (2001)]
  [hep-th/9906070].


\bibitem{Hanany:2000fq}
  A.~Hanany and B.~Kol,
  JHEP {\bf 0006} (2000) 013
  [hep-th/0003025].
  
\bibitem{Witten:1997bs}
  E.~Witten,
  JHEP {\bf 9802} (1998) 006
  [hep-th/9712028].


\bibitem{Honma:2019gzp}
  Y.~Honma and H.~Otsuka,
  arXiv:1910.10725 [hep-th].




\bibitem{Ibanez:1987pj}
  L.~E.~Ibanez, J.~Mas, H.~P.~Nilles and F.~Quevedo,
  Nucl.\ Phys.\ B {\bf 301} (1988) 157.



\bibitem{Font:1988mk}
  A.~Font, L.~E.~Ibanez and F.~Quevedo,
  Phys.\ Lett.\ B {\bf 217} (1989) 272.


\bibitem{Katsuki:1989bf}
Y.~Katsuki, Y.~Kawamura, T.~Kobayashi, N.~Ohtsubo, Y.~Ono and K.~Tanioka,
Nucl.\ Phys.\ B \textbf{341}, 611-640 (1990).



\bibitem{Kobayashi:1991rp}
T.~Kobayashi and N.~Ohtsubo,
Int.\ J.\ Mod.\ Phys.\ A \textbf{9}, 87-126 (1994).




\bibitem{Vafa:1986wx}
  C.~Vafa,
  Nucl.\ Phys.\ B {\bf 273} (1986) 592.

\bibitem{Vafa:1994rv}
  C.~Vafa and E.~Witten,
  J.\ Geom.\ Phys.\  {\bf 15} (1995) 189
  [hep-th/9409188].

\bibitem{Douglas:1998xa}
  M.~R.~Douglas,
  hep-th/9807235.
  

\bibitem{Frey:2002hf}
  A.~R.~Frey and J.~Polchinski,
  Phys.\ Rev.\ D {\bf 65} (2002) 126009
  [hep-th/0201029].
  
  




\bibitem{Cremades:2003qj}
  D.~Cremades, L.~E.~Ibanez and F.~Marchesano,
  JHEP {\bf 0307} (2003) 038
  [hep-th/0302105].



\bibitem{Blumenhagen:2005tn}
  R.~Blumenhagen, M.~Cvetic, F.~Marchesano and G.~Shiu,
  JHEP {\bf 0503} (2005) 050
  [hep-th/0502095].




\bibitem{Lust:2005dy}
  D.~Lust, S.~Reffert, W.~Schulgin and S.~Stieberger,
  Nucl.\ Phys.\ B {\bf 766} (2007) 68
  [hep-th/0506090].


\bibitem{Antoniadis:2009bg} 
  I.~Antoniadis, A.~Kumar and B.~Panda,
  Nucl.\ Phys.\ B {\bf 823}, 116 (2009)
  [arXiv:0904.0910 [hep-th]].

\bibitem{Abe:2014nla} 
  H.~Abe, T.~Kobayashi, H.~Ohki, K.~Sumita and Y.~Tatsuta,
  JHEP {\bf 1406}, 017 (2014)
  [arXiv:1404.0137 [hep-th]].

\bibitem{Cremmer:1983bf}
  E.~Cremmer, S.~Ferrara, C.~Kounnas and D.~V.~Nanopoulos,
  Phys.\ Lett.\  {\bf 133B} (1983) 61.



\end{thebibliography}
\end{document}